\documentclass{aa}

\usepackage{graphicx}

\begin{document}

\msnr{\dots}

\title{Circular Polarization of Radio Emission from Relativistic Jets}
\author{Thomas Beckert \and Heino Falcke}



\institute{Max-Planck-Institut f\"ur Radioastronomie, Auf dem H\"ugel 69, 
53121 Bonn, Germany, (tbeckert, hfalcke@mpifr-bonn.mpg.de)}

\titlerunning{Circular Polarisation in Jets}
\authorrunning{Beckert \& Falcke}

\date{submitted to Astronomy \& Astrophysics (2002)}

\abstract{
In inhomogeneous optically thick synchrotron sources a substantial
part of the electron population at low energies can be hidden by
self-absorption and overpowered by high energy electrons in optically
thin emission.  These invisible electrons produce Faraday rotation and
conversion, leaving their footprints in the linear and circular
polarized radiation of the source. An important factor is also the
magnetic field structure, which can be characterized in most cases by
a global magnetic field and a turbulent component. We present the
basic radiative transfer coefficients for polarized synchrotron
radiation and apply them to the standard jet model for relativistic
radio jets.  The model can successfully explain the unusual circular
and linear polarization of the Galactic Centre radio source Sgr A* and
its sibling M81*. It also can account for the circular polarization
found in jets of more luminous quasars and X-ray binaries. The high
ratio of circular to linear polarization requires the presence of a
significant fraction of hidden matter and low-energy electrons in
these jets. The stable handedness of circular polarization requires
stable global magnetic field components with non-vanishing magnetic
flux along the jet, while the low degree of total polarization implies
also a significant turbulent field. The most favoured magnetic field
configuration is that of a helix, while a purely toroidal field is
unable to produce significant circular polarization. If connected to
the magnetosphere of the black hole, the circular polarization and the
jet direction determine the magnetic poles of the system which is
stable over long periods of time. This may also have implications for
possible magnetic field configurations in accretion flows.
}      

\maketitle

\section{Introduction}
The detection of circular polarization (CP) in the continuum of
radio sources is believed to be a powerful tool to test
physical source models (Hodge \& Aller \cite{HA79}).
But CP in extragalactic radio sources is extremely elusive
(Roberts et al. \cite{RC75}; Ryle \& Brodie \cite{RB81};
Weiler \& de Pater \cite{WdP83}) and is detected in only a few sources.

A more recent ATCA-survey (Rayner, Norris \& Sault \cite{RNS2000}) for CP in
radio-loud Quasars, BL Lacs and Radio Galaxies with improved sensitivity of
 $0.01\%$, has shown a clear correlation of fractional CP with spectral index,
in the sense that CP is stronger in flat and inverted spectrum sources. 
Circularly polarized radiation is therefore preferentially produced in 
self-absorbed radio cores. The fractional CP at 5 GHz is found to be 
between 0.05\% and 0.5\% in 11 out of 13 inverted spectrum sources at the 
ATCA spatial resolution of $2$ arcsec. 
At higher VLBA-resolution ($\sim 0.5$mas) Homan \& Wardle (\cite{HW99})
report localized CP of
0.3\%-1\% in the jet-cores of 3C273, PKS 0528+134, and 3C279, which
in a few cases may be as high as the local linear polarization. 
It is also found, that intraday variable sources are circularly
polarized (Macquart et al. \cite{MK00}), and that LP (linear polarization)
and CP are both variable on timescales below 1 day.
Recently CP was also found in X-ray binaries (Fender et
al. \cite{Fender2000}\&\cite{Fender2002}).

While the handedness of CP is remarkably stable over several years
(Komesaroff et al. \cite{K84}; Homan \& Wardle \cite{HW99}; Fender et
al. \cite{Fender2002}) for individual sources, no preferred handedness
of CP in general is found.

An even more challenging situation than observed in radio-loud
extragalactic jet sources presents itself in the centre of our
galaxy. The compact radio source Sgr A$^*$ (see Melia \& Falcke
\cite{MF01}), believed to be coincident with the dynamical centre of
the Milky Way with a mass of $2.6\,10^6 M_\odot$ (Eckart \&
Genzel \cite{EG96}; Ghez et al. \cite{G98}) presumably in a single
black hole, exhibits consistently larger circular than linear
polarization in the range of $1.4$ to $15$ GHz (Bower et
al. \cite{BFB99}; Sault \& Macquart \cite{SM99}) with CP between
$0.2$\% and $1$\%. LP is small and below the detection limits (Bower
et al. \cite{BB99}, \cite{BW99}) with the exception of sub-mm
measurements, which possibly shows LP at a level of $10\%$ in the
range $750-2000 \mu$m (Aitken et al. \cite{A2000}). The beam size of
the sub-mm observations is $\sim 10$ arcsec.  The flux is dominated by
extended dust emission or free-free emission and the synchrotron
source is comparably weak at these wavelength.

The inverted radio spectrum of Sgr A$^*$ ($S_\nu \propto \nu^\alpha,\, \alpha
\approx 0.3$) can be interpreted as either optically thin synchrotron emission
(Beckert et al. \cite{Beckert96}) or self-absorbed synchrotron emission
from a jet-like outflow (Falcke et al. \cite{FMB93};
Falcke \& Markoff \cite{FM00}).  
The idea of synchrotron emission by thermal electrons from Sgr A$^*$ was
briefly considered by Reynolds \& McKee (\cite{RMcK80}) and revived
for mildly relativistic electrons in the
self-absorbed ADAF models for accretion in the galactic centre (Narayan et al.
\cite{N98}). The first ADAF models under-predicted the radio flux between
$1$--$100$\,GHz, which can be attributed to an outflow or jet.
The upper limits for Sgr A$^*$ in the infrared require a sharp high energy
cut-off for the electron distribution below $\gamma_{\mathrm{max}}$ of a
few\,$\times 10^2$. Therefore thermal or  quasi-monoenergetic electrons 
are responsible for the radio emission  (Beckert \& Duschl \cite{BD97}), which
distinguishes Sgr A$^*$ from high-luminosity, radio-loud AGNs.
A close relative of Sgr A$^*$ is found in the centre of the normal spiral M81.
The radio source M81$^*$ exhibits an elongated jet-like structure (Bietenholz
et al. \cite{BBR00}),
has a similiar radio spectrum (Reuter \& Lesch, 
\cite{RL96}), a slightly larger luminosity, still below the AGN level, and
has recently be found to be circularly polarized
(Brunthaler et al. \cite{Brun01}) without detectable LP.

The fractional variability of CP is usually stronger than of LP, which in turn
is stronger than for the total intensity. Together with the preserved
handedness this poses servere constrains on possible scenarios for
CP production and its variability (Komesaroff et al. \cite{K84}).
The suggested mechanisms  are ({\em a}) intrinsic
cyclo-synchrotron emission from low-energy electrons or from electrons
with small pitch angles seen close to the magnetic field direction (Legg \&
Westfold \cite{LW68}), conversion from LP to CP as a propagation effect
induced by ({\em b}) low energy electrons inside the relativistic plasma 
(Hodge \& Aller \cite{HA77}) or ({\em c}) by a magnetized cold plasma
surrounding the synchrotron source. This requires
either Faraday rotation (not possible in pure electron/positron jets)
or changing (e.g., turbulent) B-field directions along the line of sight
in the source. A further possiblity for CP production are
({\em d}) inhomogeneous rotation measures in intervening
cold plasma either close to the source or in our galaxy
(Macquart \& Melrose \cite{MM2000}).
The existence of these plasma screens can be infered from interstellar
scattering believed to be the cause for intraday variability in some 
sources (Rickett et al. \cite{R95}; Dennett-Thorpe \& de Bruyn \cite{DB00};
Macquart et al. \cite{MK00}; Beckert et al. \cite{BCore02}).
This model predicts variable CP with a time averaged mean of $<$CP$> = 0$.

In this paper we consider propagation effects like Faraday rotation
and cyclic conversion of LP to CP and back (Pacholczyk \cite{P73}) in
turbulent, self-absorbed jets or outflows. First results were already 
published in Falcke et al. (\cite{FB02}). We rederive some of the
basic radiation transfer coefficients which, for example, could also
be used for anisotropic particle distributions.  The application of
conversion to compact radio jets has been explored perviously by Jones
(\cite{J88}) using different techniques and without focusing on sources
with large circular polarization and the role of globally ordered
magnetic fields. Here we investigate the standard jet model with
respect to the new polarization data placing some emphasis on the role
of turbulence, the ratio of low- to high-energy particles, and the
magnetic field confirguation.

The paper is organized as follows: In Sec.\,\ref{emitrans} we review
the basic production channels for CP. The outfow/jet model and the
possible turbulence in the $B$-field is presented in
Sec.\,\ref{outflowmod}. The consequences of Faraday rotation and
conversion are discussed in Sec.\,\ref{dpolcon} followed by a detailed
model of Sgr A$^*$. Polarization variability is the topic of
Sec.\,\ref{polvar} and we close with a discussion of our results in
Sec.\,\ref{discus}.
\section{Polarized Synchrotron Emission \& Radiative Transfer}\label{emitrans}
\subsection{Synchrotron Emission}
Relativistic electrons or positrons gyrate in a magnetic field $B$ with
a frequency $\nu_{\mathrm{gr}} = \nu_{B \perp}/\gamma$.
The basic cyclotron frequency
\begin{equation}
  \nu_{B \perp} = \nu_B\sin\psi =
  \frac{q B \sin\psi}{2\pi m_e c} = 2.8\cdot 10^6
  \mathrm{Hz} \left[\frac{B}{1
  \mathrm{G}}\right] \sin\psi
\end{equation}
depends on the field $B_\perp = B \sin \psi$ perpendicular to the
particle momentum. The transition from cyclotron to synchrotron
radiation occurs, when emission at higher multiples of the
gyro-frequency, which are weak for non-relativistic particles, become
stronger than emission at $\nu_{\mathrm{gr}}$ due to relativistic
boosting in the instantaneous direction of the particle. The
synchrotron spectrum of individual particles reaches its maximum at
$\nu_c = 1/2 \gamma^2 \nu_B$. A population of particles with
distributions in energy and pitch angle $\psi$ produce an emission
spectrum, in which the individual cyclotron lines are blended together
and a smooth spectrum emerges. The cyclotron emission at low
frequencies is circularly polarized and reflects the spiral motion of
the particles. The emission at the maximum at $\nu_c$ is seen only for
a time $\gamma^{-2}$ of the gyration period and the weakly curved
motion during that interval produces the large linear polarization of
synchrotron emission perpendicular to the $B$-field. The emission of
individual electrons is also circularly polarized, when the angle of
the line of sight to the magnetic field $\vartheta$ is different from
the pitch angle. Relativistic beaming requires this difference $\Delta
\vartheta$ to be smaller than $\gamma^{-1}$.  The left- and
right-handed CP for different signs of $\Delta \vartheta$ nearly cancel
for an isotropic particle distribution, with a residual proportional
to $\nu^{-1/2}\,\cos \vartheta$, and cancels completely for a pure
electron-positron population. Any deviation from an isotropic pitch
angle distribution will enhance the resulting circular polarization.
\subsection{Radiative Transfer}
The transfer of polarized synchrotron radiation in homogeneous 
astrophysical plasma was derived by Sazonov (\cite{S69}) in terms
of nearly transverse electromagnetic waves (see Appendix \ref{AppA}).
The transfer equations can be
formulated for the four Stokes parameters $I,Q,U,V$ (e.g., Jones \&
O'Dell \cite{JOD77}). The effects included are emission and absorption for
$I,Q,V$ separately\footnote{A suitable transformation of coordinates leaves
$U$ parallel to the local $B$-field. Emission and absorption for $U$ vanishes
in that situation.}. The derivation of the absorption and
general rotation coefficients are given in Appendix \ref{AppB}
and \ref{AppC}. For power-law distributions of electrons (Eq. \ref{powerl}): 
$N(E) \propto \gamma^{-s}$ between $\gamma_{\mathrm{min}}$
and $\gamma_{\mathrm{max}}$, and
$\gamma_{\mathrm{min}}^2 \nu_B < \nu < \gamma_{\mathrm{max}}^2
\nu_B$, the transport coefficients are summarized in Appendix \ref{AppD}.
Here one has to be aware of the ultra-relativistic limit used in the
derivations. They are only useful for $\gamma \beta  \sin\vartheta \gg 1$.
This requirement is not always fulfilled when the power-law extends
down to $\gamma$ of a few, as will be the conclusion for Sgr A$^*$.
In addition the shape of the distribution at these energies and the assumed
perfect isotropy of the pitch-angle distribution are uncertain. The
principal treatment of more general distributions is  presented in the
Appendix.
Linear polarized emission is a fixed fraction of the total emission and
the relative emissivity of CP
(Legg \& Westfold \cite{LW68}) is
\begin{equation}
  \frac{\eta_V}{\eta_I} \propto \left(\frac{\nu}{\nu_B}\right)^{-1/2}
  \cot\vartheta \, ,
\end{equation}
where $\vartheta$ is the angle between the line of sight and the magnetic
field, which equals the pitch angle for electrons radiating into the direction
of the observer. Due to Kirchhoff's law this relation also holds for the
absorption coefficients
$\kappa_V/\kappa_I = \eta_V/\eta_I$.
Normal modes in a magnetized plasma are generally
elliptically polarized. They are circular for propagation along the
magnetic field and coupling to the gyration of electrons induce different
refractive indices for left- and right-handed modes. This leads to Faraday
rotation described by the transfer coefficient $\kappa_F$, which has the
dimension of an absorption coefficient, so that after propagating
a path length $\Delta l$ linear modes rotate by an angle
$\Delta \zeta = \Delta l\,\kappa_F$.

Due to the steep frequency dependence of all rotation coefficients
Faraday rotation is important at low frequencies 
\begin{equation} \label{coeffF}
  \kappa_F \propto \left(\frac{\nu}{\nu_B}\right)^{-2}
  \frac{\ln \gamma_{\mathrm{min}}}{\gamma_{\mathrm{min}}^{s+1}}
  \cos\vartheta \quad,
\end{equation}
with the power-law index $s$ for the electrons. Intrinsic Faraday rotation
is dominated by the low-energy end of the electron population and
depends on the line of sight average parallel magnetic field in the
source.

Perpendicular to the field the normal modes are linearly polarized
and different refractive indices for modes parallel and perpendicular
to the $B$-field leads to bi-refringence and cyclic transformation of
$U$ into $V$ with a rotation angle $\sin \Delta \zeta' = \Delta l\,\kappa_C$ 
when propagating a distance $\Delta l$. This effect was termed
'repolarization' by Pacholczyk (\cite{P73}) and is described by the transfer
coefficient $\kappa_C$ (see Appendix \ref{AppD})
\begin{equation}
 \kappa_C \propto  \left(\frac{\nu}{\nu_B}\right)^{-3} \sin^2\theta
  \gamma_{\mathrm{min}}^{-(s-2)} \, ,
\end{equation}
if the source is self-absorbed and emission at frequency $\nu$ is dominated
by high-energy electrons with $\gamma_{\mathrm{rad}} \gg \gamma_{\mathrm{min}}$
or equivalently $\nu \gg \nu_B \gamma_{\mathrm{min}}^2$.
If we have additional cold plasma in or surrounding the
emission region, the radiation is modified along its path by additional
Faraday rotation and conversion $\kappa_F^{(c)}$ and $\kappa_C^{(c)}$
(see Appendix \ref{AppC}). It must be noted that no substantial amount of
cold plasma $n_{(c)} \approx n_{(r)}$ compared to the density of
relativistic particles can exist in linearly polarized, compact synchrotron
sources ($T_B \sim 10^{10\ldots 12}$\,K) without depolarizing the emission
(Jones \& O'Dell \cite{JD77}). 
Inside a source with a homogeneous and well ordered magnetic field Faraday
rotation is necessary for conversion, because no linear polarized
component $U$ with electric field parallel to the stationary B-Field is
otherwise produced.

An electron-positron plasma on the other hand shows no Faraday rotation and
changing magnetic field directions along the line of sight either in an
ordered field structure and/or in a turbulent field is required to start
conversion. The contribution of protons to Faraday rotation 
and conversion is weaker by a factor $(m_e/m_p)^3$ and
$(m_e/m_p)^2$ respectively. In view of the $\gamma_{\mathrm{min}}$
dependence in Eq. (\ref{coeffF}) this implies that Faraday rotation by
cold protons is as important as by electrons, if
$\gamma_{\mathrm{min}}^{2s}/(\ln \gamma_{\mathrm{min}}) \sim 3/4(m_p/m_e)^3$.
So for $s=2$ a $\gamma_{\mathrm{min}} \sim 475$ corresponds
to equally strong rotation by electrons and protons and consequently
no rotation. This is not a strong constraint, but for $s=3$ we get equally
strong rotation from cold protons and relativistic electrons
for $\gamma_{\mathrm{min}} = 50$ and for larger $\gamma_{\mathrm{min}}$
the protons dominate rotation. 
\subsection{Intuitive Approach}
To visualise the effects and properties of radiation transfer in a
compact medium and achieve a more intuitive understanding of the
relevant effects on polarization, we also provide a more qualitative
discussion here.  For simplicity let us separate Faraday rotation from
conversion and only picture purely linearly or circularly polarized
waves in a homogeneous magnetic field.

\begin{figure}
    \resizebox{\hsize}{!}{\includegraphics{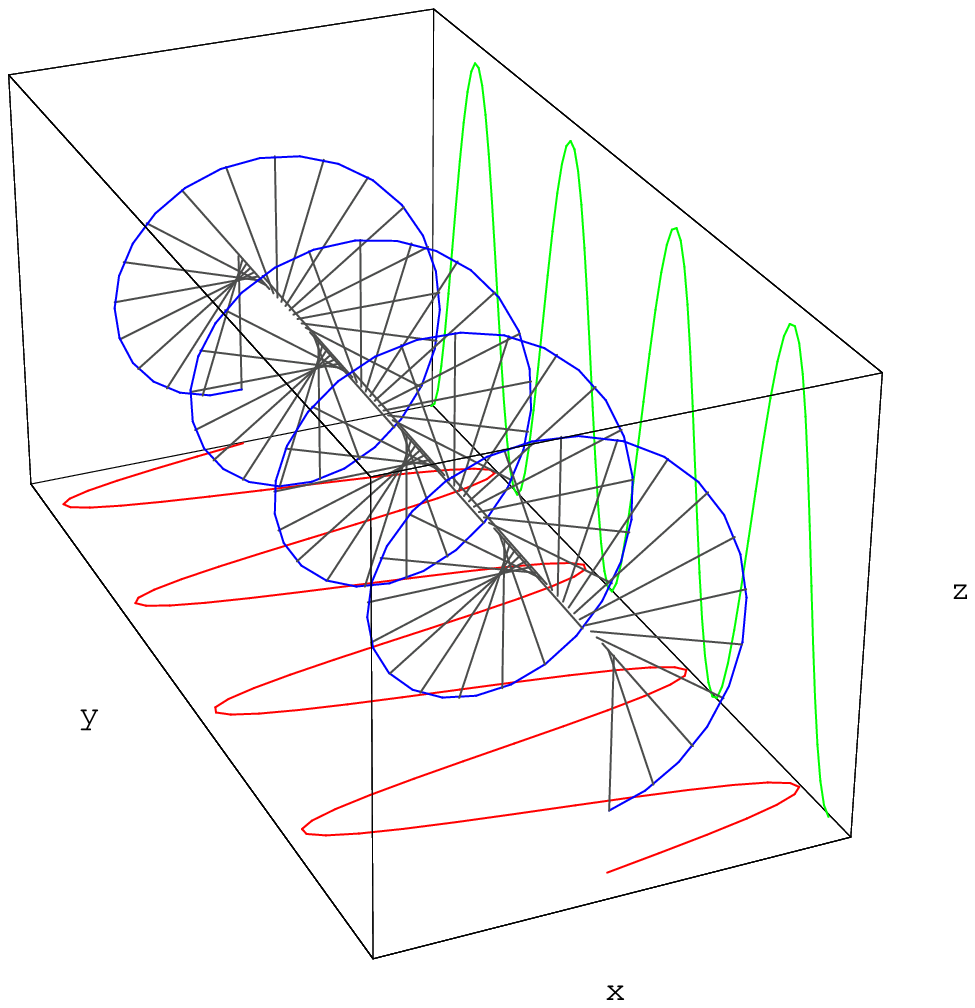}}
    \caption{A circularly polarized wave can be composed of two
    orthogonal linearly polarized modes shifted in phase. A phase
    shift would be produced by a plasma in a magnetic field
    perpendicular to the propagation direction of the waves (here
    along the $z$-direction). Without phase-shift the sum of the two
    modes would be a purely linearly polarized wave.  The accompanying
    movie shows the effect of how phase-shifts in a region
    will turn such a linearly polarized wave in to a circularly
    polarized wave (conversion).}
    \label{conv-schematic}
\end{figure}

The two orthogonal normal modes for propagation perpendicular to the
magnetic field are linearly polarized and a purely circularly
polarized wave is split into the two normal modes with a relative
phase shift as shown in Fig.~\ref{conv-schematic}. Without a
phase-shift the wave will be purely linearly polarized.  If, for
example, a locally homogeneous magnetic field vertically pervades the
box in Fig.~\ref{conv-schematic} along the $z$-direction, electrons or
positrons will be free to move along the field lines and resonate with
the vertical mode but hardly resonate with the horizontal mode along
the $x$-direction. This yields the bi-refringence discussed above. The
resonating electrons or positrons will themselves act as antennas and
emit a somewhat delayed wave that interferes with the incoming
vertical mode, leading to a slight phase-shift between vertical and
horizontal mode. The effect of this shift is shown in the accompanying
animation\footnote{See also the authors webpage at
http://www.mpifr-bonn.mpg.de/staff/hfalcke/CP} of
Fig.~\ref{conv-schematic}, where the resulting wave is circularly
polarized and switches from linear to circular polarization as a
function of the shift. 

Conversion acts also on initially only linearly polarized radiation.
The amount of this conversion will depend on the misalignment between
the incoming wave and the magnetic field direction since, obviously, a
phase-shift between two orthogonal modes will have little effect if
one mode is very small or non-existent. Moreover, a random
distribution of magnetic field lines on the plane of the sky will
reduce circular polarization from conversion in exactly the same way
as linear polarization would be reduced.

\begin{figure}
    \resizebox{\hsize}{!}{\includegraphics{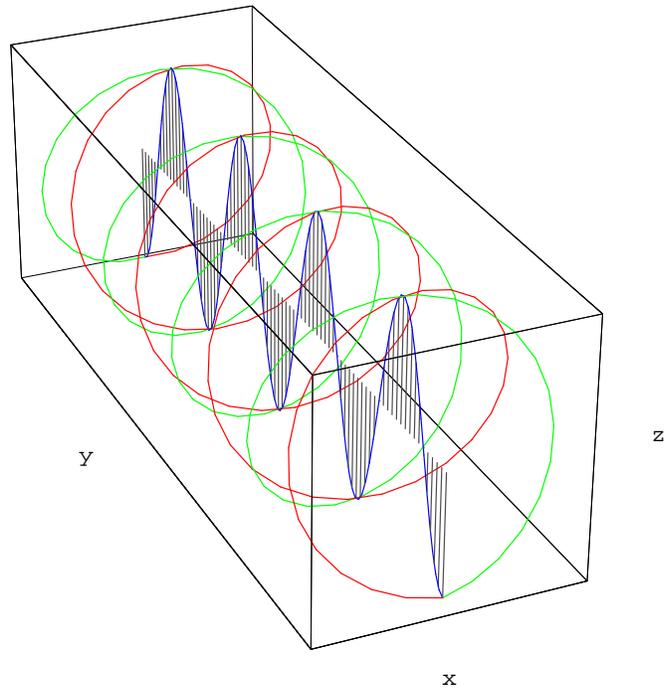}}
    \caption{A linearly polarized wave can be composed of two
    orthogonal circularly polarized modes shifted in phase. A phase
    shift would be produced by a plasma in a magnetic field along the
    propagation direction of the waves (here along the
    $y$-direction). The accompanying movie shows the effect of
    additional phase-shifts on the linear polarization, leading to
    Faraday rotation.}
    \label{faraday-schematic}
\end{figure}

Analogous to the picture for conversion, one can view a linearly
polarized wave as composed of two circularly polarized normal modes when
propagating along the magnetic field. This is
sketched in Fig.~\ref{faraday-schematic}, where we will assume a
longitudinal magnetic field, i.e. a field along the $y$-direction. The
circular modes will resonate with either electrons or positrons
gyrating around the magnetic fields. The latter will again emit a
circularly polarized wave, producing a phase-shift when interfering
with the incoming wave. The effect of the phase-shift in the circular
modes is shown in the accompanying animation of
Fig.~\ref{faraday-schematic}$^2$, where one can see that the resulting
linearly polarized wave is simply (Faraday) rotated.

An important conclusion to remember therefore is, that conversion is
mainly produced by magnetic field components perpendicular to the
line-of-sight or photon direction, while Faraday rotation is produced
by magnetic field components along the line-of-sight. Moreover, one
can also see that conversion is insensitive to the electron/positron
ratio while Faraday rotation is not. In Fig.~\ref{conv-schematic} an
electron and an positron are both free to move along the
$z$-axis. While they will respond in opposite directions to the
incoming wave, their respective emitted waves will also have opposite
signs because of opposite charges and hence be identical. In the case
of Faraday rotation, the incoming left- or right-handed circularly
polarized wave will only resonate with the particle that also has the
correct handedness in its gyration -- either electron or positron
depending on the magnetic field polarity. A pure pair plasma would
therefore produce exactly the same phase shift in left- and
right-handed modes and not produce any net Faraday rotation. In the
case of a charge-excess, the direction of Faraday rotation depends on
the sign of the charge-excess (presumably electrons) and the polarity
of the magnetic field. This will indirectly also affect the sign of
circular polarization, if Faraday rotation is the ultimate cause of
the misalignment between the plane of polarization and the magnetic
field direction.

\section{Outflow/Jet Models}\label{outflowmod}
\subsection{Conical Outflows} \label{outflow}
Models of flat spectrum radio cores in AGN assume in general a conical
jet (Blandford \& K\"onigl \cite{BK79}; Falcke \& Biermann
\cite{FB95}), in which plasma is flowing out with constant velocity $v
= \beta c$ and constant half opening angle $\theta = \arcsin(R/z)$,
where $z$ is the coordinate along the jet and $R$ the local radius of
the jet.  The magnetic field in the jet must have an ordered
component, which leads to persistent polarization, and probably a
turbulent field\footnote{Alfvenic turbulence is expected from
diffusive particle acceleration at shock fronts.  Pitch angle
scattering at Alfven waves will also lead to isotropic particle
distributions, which we assume here.}. The ordered large-scale field
can be separated in modes, which carry magnetic flux and will
therefore decay as $\sim z^{-2}$, and modes without magnetic flux
(e.g. toroidal fields), which behave as
\begin{equation}
  B = B_0 \left(\frac{z}{z_0}\right)^{-1} \qquad .
\end{equation}
The second type of modes dominate at large radii.
One possible configuration is a force-free
equilibrium for a parallel jet (K\"onigl \&  Choudhuri \cite{KC85}).
The resulting structure can show both a mode with magnetic flux and modes
of twisted flux tubes with toroidal and
two oppositely directed poloidal fields, so that the magnetic flux along $z$
vanishes for the second modes. Other configuration may not be force-free
but are equally valid. The most likely structure is a helical $B$-field due
to the rotation of the footpoints in an accretion disk. The spiral can be
regarded as the superposition of a field component $B_z$ along $z$ and a
toroidal component $B_\phi$. The slope of the spiral is
$\alpha_S = \arcsin(B_\phi/B_z)$. 

Every jet in perpendicular pressure equilibrium ($r$-direction) with 
its surrounding will suffer adiabatic energy losses. Relativistic electrons,
which are injected at the base of the jet, will cool down due to adiabatic
expansion, which leads to inverted radio spectra observed in some core
dominated
extragalactic radio sources. Further along the jet electrons have to be
reaccelerated. Alternatively the jet is highly over-pressured relative to
its surrounding and adiabatic losses are negligible consistent with flat
radio spectra for conical jets cores.
It should be noticed, that toroidal
fields will induce electric currents and hook stresses will confine the jet,
if the magnetic field is well ordered and strong enough to influence the 
jet dynamics.
Otherwise a cosmic conspiracy of electron cooling (adiabatic and radiative
losses) and geometric changes, which includes the evolution of magnetic 
fields, must be proposed. The variety of observed spectral indices between 
1 and 20 GHz indicates, that intermediate stages, where several of the 
effects are present, are quite common.
The particle distribution without pair-production $N(E,z)$ in a cone can
vary as
\begin{equation}
  N(E,z) = N(E,z_{\mathrm{in}}) \left(\frac{z}{z_{\mathrm{in}}}
  \right)^{-2(1+ a/3)}
\end{equation}
where $N(E,z_{\mathrm{in}}) = N(E)$ is the energy distribution at the
injection point. The parameter  $a$ is zero for a
freely expanding and over-pressured jet and $a=1$ for adiabatic losses
in pressure equilibrium. Radiative cooling can further change the distribution
and may be considered separately.
For a conical outflow with partial adiabatic particle cooling the resulting
sychrotron spectrum can be separated in three spectral regimes.
\begin{equation}
 S_\nu \propto \nu^\alpha\,, \quad \left\{ \begin{array}{cc}
            \alpha = 5/2 & \mbox{optically thick}\\
            \alpha = \frac{5}{2}\left(1 - \frac{s + 4}{s + 4 +4 a/3}\right) &
            \mbox{self-absorbed} \\
            \alpha = (s-1)/2 & \mbox{ optically thin}
            \end{array} \right.
\end{equation}
The optically thick regime can only be observed if the outflow terminates
at a maximal distance, or fragments into subcomponents, which break the
self-similarity of the conical model.
In the self-absorbed regime, the flux is dominated by emission from the
region around the optical surface ($\tau$=1-surface). A
sufficiently large ratio $z_{\mathrm{out}}/z_{\mathrm{in}}$
of outer to inner radius of the self-similar conical outflow is
necessary for it to be observable. Optically thin emission is always
present at frequencies above $\nu_{\mathrm{in}} =
\nu_{\mathrm{abs}}(z_{\mathrm{in}})$, if not other radiation
processes like bremsstrahlung or dust emission dominate. In some sources the
low-frequency spectrum can change due to free-free-absorption.
\subsection{Turbulence in Jets}
Synchrotron emission from self-absorbed radio sources with brightness
temperatures $T_B$
of $\sim 10^{11}$K imply near energy equipartition of radiating
electrons and/or positrons and magnetic field (Readhead \cite{R94}).
This finding is modified by relativistic Doppler boosting for
variable flat spectrum radio cores. VSOP-observations show that the
observed $T_B$ can be larger than $10^{12}$K (Bower \& Backer \cite{Bow98};
Tingay et al. \cite{T01}) in selected sources.
This agrees with the observed superluminal speeds
$\beta_{\mathrm{obs}} \sim 5$, which require Doppler factors $ > 5$.
The fractional LP of radio cores is usually only a few per cent or smaller
and requires either strong Faraday depolarization (Tribble \cite{T91})
or tangled $B-$fields. The observed rotation 
measures in some quasars (Taylor \cite{T00}) are not sufficient
to depolarize the radio emission at cm-wavelength\footnote{This is not
true anymore, if the rotation measure rises sharply 
towards the centre ($RM \propto z^{-2}$ or steeper).} and we infer the
presence of a turbulent contribution to the global $B$-field described above.
The turbulent field is effectively depolarizing the source, if the
amplitude of the dominating Fourier modes of the turbulent field are about
a factor of 2 larger than
the local contribution of the global field, so that local field reversals
occur.
We describe the turbulent magnetic field as a local
superposition of incoherent waves with wavenumber $k$, which decorrelate
over distances of a few times their wavelength. For the treatment of radiative
transfer in turbulent jets, we consider the turbulence to be frozen in and
time evolution to be unimportant. 

The turbulent wave spectrum is characterized by an outer wavelength 
and corresponding wavenumber $k_{\mathrm{out}} = 2\pi/L$ and a
dissipation wave number $k_d$. Between these wave numbers an inertial
range with energy cascading 
from small to large wave numbers will develop\footnote{Here it is implied,
that an  instability exists, which injects energy into the turbulent
cascade at the outer wavelength.}.
The local strength and orientation  of the turbulent magnetic field will
be determined by
modes with wave numbers around $k_{\mathrm{out}}$ and the typical
length-scale for changes of the magnetic field is
$|B/\nabla B| \sim k_{\mathrm{out}}^{-1}$. 

We assume that the spectral energy density in the inertial range is
described by a Kolmogorov spectrum
\begin{equation}
  E(k) = F\ \epsilon^{2/3}\ k^{-5/3} \quad .
\end{equation}
Here $F$ is a number of order unity.
The dissipation wave number is related to the energy dissipation rate 
$\epsilon$ by the inequality
(e.g. Frisch \cite{F95})
\begin{equation} \label{dissl}
   k_d^{4} \ge \epsilon\ \nu^{-3} 
\end{equation}
where $\nu$ is the viscosity.
The dissipation $\epsilon$  for Alfvenic turbulence is then given by 
\begin{equation} \label{Edisp}
    \epsilon = \left(B_T^2/(B_0^2 F)\right)^{3/2}\,k_{\mathrm{out}}\,v_A^3\ ,
   \qquad
    v_A^2 = \frac{B_0^2}{8\pi \rho} \, ,
\end{equation} 
with $v_A$ the Alfven velocity in the global field and $B_T^2/B_0^2$ the
energy density ratio of turbulent to the global field. We can combine the 
dissipation rate (\ref{Edisp}) with the estimate for the dissipation wave
number (\ref{dissl}) to get the viscosity
\begin{equation} \label{visco}
  \nu \ge (B_T^2/B_0^2 F)^{1/2} (k_{\mathrm{out}}/k_d)^{1/3}\,v_A\,k_d^{-1}\,.
\end{equation}
If the energy dissipation process is known, it is possible to estimate the
thermalisation of magnetic energy along the jet. One possibility is a first
resonance with the gyration of protons or relativistic electrons. 
\subsection{Error Estimate for Radiative Transfer in Turbulent Fields} 
A numerical treatment of radiative transfer in turbulent magnetic field will
not resolve incoherent fluctuations with wave numbers larger than $k_n$ 
depending on the numerical code. The largest effect possible of the
unresolved modes  with $k_n < k < k_d$ is a coherent addition.
The critical wave number $k_c$, for which the rest $k > k_c$ can result
in field reversals is 
\begin{equation}
    k_c = k_{\mathrm{out}} \left[ \frac{k_{\mathrm{out}}
    B_{k_{\mathrm{out}}}}{B_0}\frac{1}{6 <\cos \phi_k>_k} + 1
    \right]^{6}. 
\end{equation} 
We expect to overestimate emission and absorption of polarized radiation, 
which depends on the projection of the magnetic field onto the line of sight 
and which change sign with field reversals. Using the estimate of 
$<\cos \phi_k>_k \approx 2/\pi$, we have an upper limit to the reduction of
the effective absorption and emission coefficients by
\begin{equation}
    p = \frac{\Delta B_n}{\Delta B_c} \approx 
   \frac{2}{\pi}\frac{6 B_0}{k_{\mathrm{out}} B_{k_0}}
   \left[\left(\frac{k_n}{k_0}
   \right)^{1/6} - 1\right] \quad .
\end{equation} 
In the numerical models the smallest length-scale $k_n^{-1}$ is adjusted, so
that $p > 1$ and no reduction has to be applied.

\section{Depolarization and Conversion}\label{dpolcon}
\begin{figure}
    \resizebox{\hsize}{!}{\includegraphics{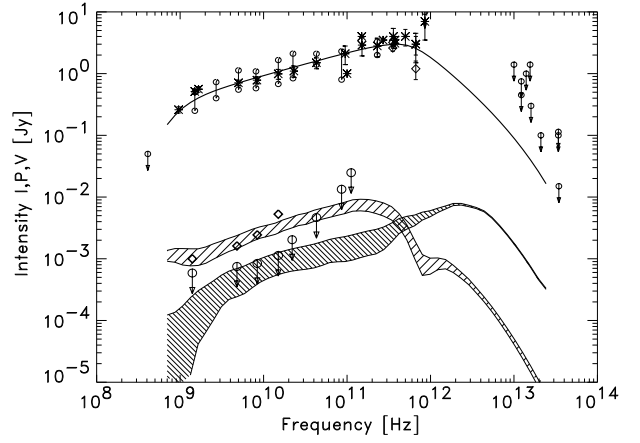}}
   \caption{Outflow model for the radio spectrum of Sgr A$^*$.
   The result of model calculations for total flux $I$
   (solid line), linear (dense shaded area)
   and circular polarized flux $V$ (sparse area) are shown for a distance
   of $8$ kpc. The shaded areas mark the expected variability due to
   turbulence with $k_{\mathrm{out}} R = 50$. The global magnetic field
   structure is a
   spiral with $B_\phi/B_z = 1$. CP data are taken from Bower (\cite{Bow00}),
   averaged from two campaigns. LP measurements are from Bower et al.
   (\cite{BW99}) and (\cite{Bow01}) }
   \label{sgrAfig}
\end{figure}
\begin{figure}
    \resizebox{\hsize}{!}{\includegraphics{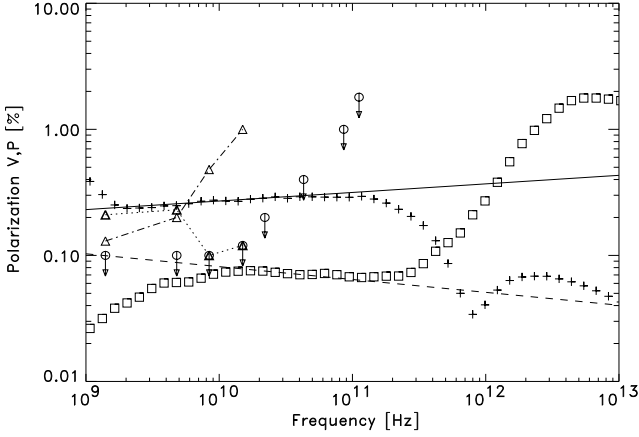}}
  \caption{Fractional polarization for the numerical calculations ($\pi_L$:
  boxes; $\pi_C$: crosses)
  shown in Fig.\ref{sgrAfig} compared to the upper limits for linear (arrows)
  and the measured circular polarization (triangles).
  Data are taken from two campaigns
  (Bower \cite{Bow00}).
  The Percentage of linear polarized flux (solid line) and circular
  (dash-dotted)
  for the analytic estimates of Eq. (\ref{linFarad}) and Eq. (\ref{circFard})
  is also shown.}
   \label{sgrAPol}
\end{figure}
\subsection{Analytical Estimates}
Turbulence leads to a reduction of measured polarized flux, if observations
are not able to resolve the largest turbulent scales $L \approx
k_{\mathrm{out}}^{-1}$ in the flow.
Additional reduction of polarization occurs along the ray path
through the source, if the Faraday optical depth in regions of size $L$
is large. We define the Faraday cell depth as 
\begin{equation} \label{tauFk}
  \tau_F = \kappa_F/k_{\mathrm{out}} \, ,
\end{equation}
where $\kappa_F$ is the Faraday rotation coefficient given in (\ref{Frotate}).
At a given frequency only the polarized flux from the optical surface
is important and will be considered, even if the fractional polarization in
the optically thin region above the optical surface is larger.
Both conversion and rotation changes the linear polarized flux and
their relative importance is measured by
\begin{equation} 
   \xi = -\frac{\kappa_F}{\kappa_C} = \frac{s + 2}{s + 1}\frac{\nu}{\nu_B}
   \frac{\cos\vartheta}{\sin^2\vartheta}
    \frac{\ln \gamma_{\mathrm{min}}}{\gamma_{\mathrm{min}}^3}
    \frac{s - 2}{1 -
    \left(\frac{\gamma_{\mathrm{min}}^2 \nu_B}{\nu}\right)^{s/2-1}}
    \, .
\end{equation}
For flat spectrum self-absorbed outflows the ratio
$\nu/\nu_B = \gamma_{\mathrm{rad}}^2$ is
nearly constant and radiation is dominated by electrons with the same
$\gamma_{\mathrm{rad}}$ at all radii. 
For the special case $s = 2$ we get
\begin{equation} 
   \xi = \frac{4}{3} \frac{\cos\vartheta}{\sin^2\vartheta}
       \left(\frac{\gamma_{\mathrm{rad}}}{\gamma_{\mathrm{min}}}\right)^2
       \frac{\ln \gamma_{\mathrm{min}}}{\gamma_{\mathrm{min}}
       \ln(\gamma_{\mathrm{rad}}/\gamma_{\mathrm{min}})}  \, . 
\end{equation}
\begin{figure}
    \resizebox{\hsize}{!}{\includegraphics{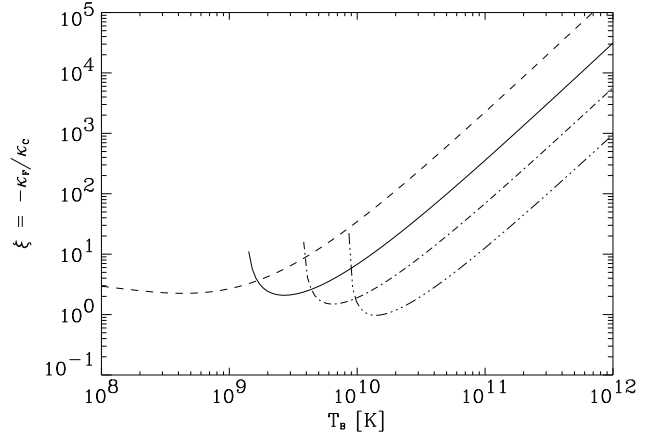}}
  \caption{The ratio $\xi$ of Faraday rotation to conversion as function
           of brightness temperature for the
           intermediate angle $\vartheta_0$, defined by
           $\cos\vartheta_0 = \sin^2\vartheta_0$. The electron spectral
           index is $s=2.5$ and $\xi$ is plotted for four values of
           $\gamma_{\mathrm{min}} = $2 (dashed line), 5 (solid line),
           10 (dash-dotted), and 20 (dashed-3$\times$dotted). }
   \label{xiRatio}
\end{figure}
The dependence of $\xi$ on the viewing angle $\vartheta$ is $\xi
\propto \cos\vartheta/\sin^2\vartheta$ and the ratio $\xi$ is shown in
Fig.\,\ref{xiRatio} as a function of brightness temperature $T_B
\propto \gamma_{\mathrm{rad}}$ for different values of
$\gamma_{\mathrm{min}}$.  Faraday rotation is less important than
conversion, if $\gamma_{\mathrm{rad}} > 20$ and $\gamma_{\mathrm{min}}
\approx 0.9\gamma_{\mathrm{rad}}$.  Whenever the power-law population
of electrons extends below $\gamma_{\mathrm{min}} = 20$, Faraday
rotation is always stronger than conversion with the exception of
almost perpendicular magnetic fields.  For high brightness
temperatures and $\gamma_{\mathrm{min}}< 0.5\gamma_{\mathrm{rad}}$
Faraday rotation depolarizes the emission, which is applicable to the
situation in Sgr A* discussed below.

The polarization is determined close to the optical surface and the
relative Faraday optical depth is
\begin{equation} \label{tauF} 
   \tau_F/\tau \approx \frac{2(s+2)}{(s+1) k_{\mathrm{out}} R}
   \left(\frac{\gamma_{\mathrm{rad}}}{\gamma_{\mathrm{min}}}\right)^s
   \frac{\ln\gamma_{\mathrm{min}}}{\gamma_{\mathrm{min}}} \, ,
\end{equation}
where $k_{\mathrm{out}} R$ is the number of turbulent cells along the ray path.
In analogy the relative strength of conversion is
\begin{equation} \label{tauQ}
   \tau_C/\tau \approx 
   \frac{2}{(s-2)\,k_{\mathrm{out}} R}\left[\left(\frac{\gamma_{\mathrm{rad}}}
   {\gamma_{\mathrm{min}}}\right)^{s-2}-1\right]
\end{equation} 
Consequently for $\gamma_{\mathrm{rad}} \approx 100$ substantial Faraday
rotation occurs for
$\gamma_{\mathrm{min}}$ less than $10$.

In sources where the relative Faraday optical depth (\ref{tauF}) is larger
than unity the depolarization along a ray path is dominated by Faraday
rotation within cells of size $k_{\mathrm{out}}^{-1}$. The optical surface
of the source is covered by $(k_{\mathrm{out}} R)^2$ cells. 
The instantaneous fractional LP
due to internal Faraday rotation is then 
\begin{equation} \label{linFarad}
  \pi_L = \frac{\sqrt{Q^2 + U^2}}{I} \approx
   \frac{s + 1}{s + 7/3}\frac{1}{k_{\mathrm{out}} R}
   \tau/\tau_F \, .
\end{equation}
The time average $\left<\pi_L\right>$ will
vanish in a stochastic magnetic field. In jets or outflows from rotating
central objects we expect an ordered magnetic field component. For the 
spiral magnetic field described in Sec.\ref{outflow} the averaged field
in the plane of the sky is $B_z \sin\vartheta$ and we expect a mean
fractional LP of
\begin{equation} \label{linFmean}
\left<\pi_L\right> \approx
   \frac{s + 1}{s + 7/3}\frac{1}{k_{\mathrm{out}} R}
   \frac{\tau}{\tau_F} \frac{B_z}{B_0} \sin\theta\, .
\end{equation}
The field strength $B_0$ is the sum of globally ordered and turbulent
field and $\theta$ the angle between the jet direction and the line of sight.

The appearance of linear polarized radiation perpendicular to the
local magnetic field is the starting point for conversion to circular
polarized radiation. Linear polarized radiation from outside the cell will
have a non-vanishing Stokes U locally, while LP emission intrinsic to the
cell must undergo Faraday rotation before conversion can take place.
In the case of dominating Faraday rotation $\tau_F > \tau$
the fraction of suitable LP along a particular line of sight is
$(p + 1)/(p + 7/3) \tau/\tau_F$. This fraction can
be converted with an efficiency $\tau_C/\tau$, if $\tau_C < \tau$. The
resulting fractional CP follows from an average over surface elements
of relative size $1/(k_{\mathrm{out}} R)^2$: 
\begin{equation} \label{circFard}
  \pi_C = \frac{\sqrt{V^2}}{I} \approx \frac{s + 1}{s + 7/3}
   \left(\frac{1}{k_{\mathrm{out}} R}\right)
   (\tau/\tau_F)(\tau_C/\tau)\,.
\end{equation}
Again the time average CP in a stochastic field will vanish
$\left<\pi_C\right>=0$.
In an outflow with a spiral magnetic field component the average CP
is approximately 
\begin{equation} \label{circFmean}
  \left<\pi_C\right> \approx 0.5 \frac{s + 1}{s + 7/3}
   \left(\frac{1}{k_{\mathrm{out}} R}\right)
   \frac{\tau_C}{\tau_F}\,\frac{B_z}{B_0} \cos\theta\,.
\end{equation}
If Faraday rotation within one cell is small, the relevant perpendicular
polarized emission is proportional to $(k_{\mathrm{out}} R)^{-1/2}$
and the resulting CP
\begin{equation} 
  \pi_C \approx
   \left(\frac{1}{k_{\mathrm{out}} R}\right)^{3/2}
   (\tau_C/\tau) \quad .
\end{equation}
In that case LP is not reduced by Faraday depolarization along a
particular line of sight and
\begin{equation} 
  \pi_L  \approx
   \frac{p + 1}{p + 7/3}\left(\frac{1}{k_{\mathrm{out}} R}\right)^{3/2}
\end{equation}
the resulting linear polarized flux is larger than the circular
by a factor $\tau/\tau_C$, which is larger than one.
\subsection{The Specific Model for Sgr A$^*$}
\begin{figure}
    \resizebox{\hsize}{!}{\includegraphics{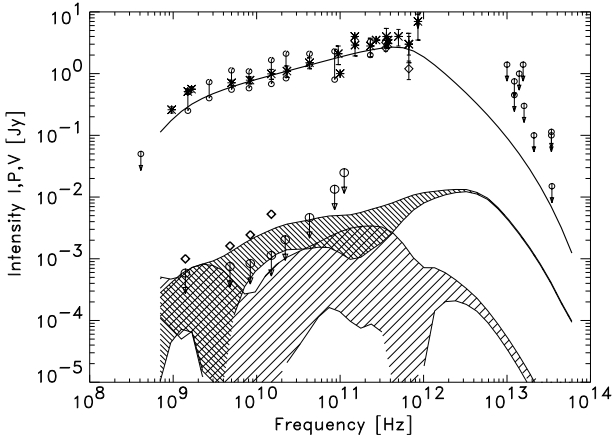}}
  \caption{The same model for Sgr A$^*$ as shown in Fig. 1 with a tightly
   wound spiral structure $B_\phi/B_z = 3$ seen at an angle of $30^\circ$
   to the jet axis. It can not explain the observed CP.}
   \label{sgrAPol_A}
\end{figure}
The inverted spectrum of Sgr A$^*$ implies either a non-conical
geometry of the outflow, a magnetic field component, which decays
faster than $1/z$ along the outflow, or reduced, but not absent,
adiabatic cooling of the radiating, relativistic
particles. Acceleration of the ouflowing plasma (Falcke \cite{F96})
also produces inverted radio spectra. For simplicity, this is not
explicitly considered here, however, its effect is essentially covered
by the assumed scaling of magnetic field and density.  Cooling of
relativistic particles implies $N(\gamma, z) \propto
\gamma^{-s} z^{-2(1+a/3)}$, where $a=0$ for freely expanding outflows,
and $a=1$ for adiabatic losses in pressure equilibrium with the surrounding
gas. The spectral index $\alpha = 0.3$ implies $a = 0.1 (s+4) \approx 0.7$
for our choice of the electron spectrum $s=3$. This is one of the possible
parametrisations of the observed inverted spectrum. The magnetic field at the
base of the outflow is $60$\,G at $z_{\mathrm{in}} = 10\,R_S$ in our model,
where the width of the outflow equals the distance from the black hole of
mass $2.6\,10^6\,M_\odot$.
The presumed electron distribution extends
from $\gamma_{\mathrm{min}} = 5$ to $\gamma_{\mathrm{max}} = 250$.
The high energy cut-off is required by the infrared limits of Sgr A$^*$
spectrum and resembles a wide
quasi-monoenergetic distribution (Beckert \& Duschl 1997). The low-energy
end is set by Faraday rotation and conversion to produce the observed
polarization in an inhomogeneous, optically thick jet or outflow. The emission
becomes optically thin above $\nu = 5\cdot 10^{11}$ Hz, which implies
that optical depth is unity for $\gamma_{\mathrm{rad}}
\approx 35$ at the base of the outflow.
Due to the inverted spectrum $\gamma_{\mathrm{rad}}$ varies with radius
$\gamma_{\mathrm{rad}} \propto z^{-0.068}$, which has to be considered in Eq.
\ref{tauF} and Eq. \ref{tauQ}. This implies a weak
frequency dependence of $\left<\pi_L\right>$ and $\left<\pi_C\right>$ of the
model seen in Fig \ref{sgrAPol}.
Because Faraday rotation is strong and conversion within one cell  
is weak in our model with $\tau_H/\tau \sim 6.04/(k_{\mathrm{out}} R)$, where
we use an outer scale $k_{\mathrm{out}} = 50/R$, we can
use Eq. (\ref{circFmean}) and Eq. (\ref{linFmean}) to estimate the
mean CP and LP. The analytical estimates for $\left<\pi_L\right>$ and
$\left<\pi_C\right>$ are
shown in Fig. \ref{sgrAPol} together with the measured CP and the results
from numerical solutions of the transfer problem.
These estimates hold as long as the outflow is self-absorbed below
$3\cdot 10^{11}$Hz.

The model for Sgr A$^*$ requires depolarization dominated by a large
Faraday rotation depth. In this case the outer turbulent scale
$k_{\mathrm{out}}$ is poorly constraint. For the analytic model of
Eq. (\ref{linFmean}) and (\ref{circFmean}) to be valid  
$k_{\mathrm{out}}$ has to satisfy $6 \ll k_{\mathrm{out}} R \ll 1250$.
The lower bound comes from $\tau_C/\tau \ll 1$ and the upper bound
from $\tau_F/\tau \gg 1$. The value for the numerical
treatment shown in Fig.\,\ref{sgrAfig}
and Fig.\,\ref{sgrAPol} is $k_{\mathrm{out}} R = 50$.
The resulting spectra for numerical solutions of the radiative transfer
problem (Eq. \ref{TransPro}-\ref{TransProEnd}) on a $100^2$ grid
covering the jet seen under
an angle of $\theta = 30^\circ$ in the rest frame of the gas is shown
in Fig. \ref{sgrAfig}.
The required electron density is
$2.8\,10^7 \left(z/10\,R_S\right)^{-2}$\,cm$^{-3}$ in a global
$B$-field of $60 \left(z/10\,R_S\right)^{-1}$\,G.
The half opening angle of the subsonic outflow is $4.5^\circ$ with a
bulk motion of $\beta = 0.4$. It turns out that a spiral structure for the
global magnetic field seen under an angle $\theta < \pi/2 -\alpha_S$ is
preferred for reproducing the level of linear and circular polarization. The
numerical simulation use a $\alpha_S=\pi/4$-spiral corresponding to
$B_z/B_0 = 1/\sqrt{2}$. 
In the limit of a very long spiral with $\alpha_S \rightarrow 0$ the
jet must be
seen at angle smaller than $\sim 40^\circ$, because conversion and Faraday
depolarization is stronger for small $\theta$ in our model. For the
assumed electron-proton plasma the kinetic bulk energy is about
$7.4\,10^{39}$\,erg/s. Together with the magnetic energy flux, the
thermal energy flux derived from the half opening angle of the flow,
and the energy supply needed to overcome the gravitational potential with
the large mass loading of the flow
starting at $10\,R_S$, the total power is $L_{\mathrm{Jet}}
\approx 1.3\,10^{40}$\,erg/s. From Eq. (\ref{Edisp}) we can estimate
the turbulent energy dissipation rate
along the jet to be $ 2.7\,10^{39}  
\ln (z_{\mathrm{out}}/z_{\mathrm{in}})[(k_{\mathrm{out}} R)/50]$\,erg/s.
The inverted part of the spectrum of Sgr A$^*$ extends from 1\,GHz to
350\,GHz and the ratio $z_{\mathrm{out}}/z_{\mathrm{in}}$ must
therefore be larger 350. This provides a upper bound to the outer
turbulent scale $k_{\mathrm{out}} R < 24$ so that the kinetic energy
is not dissipated before reaching $z_{\mathrm{out}}$. The total jet
power is 5 orders of magnitude larger than the emitted radio
luminosity, which increases the required accretion rate to power the
jet. The brightness temperature of the model is little less than the
equipatitition temperature and the inverse Compton luminosity from
optical to X-rays is much less than the radio luminosity. The radio
jet of Sgr A$^*$ is a very inefficient radiation source. For the
polarization in M81* we would qualitatively obtain similar numbers,
with however, a higher jet power.

\subsection{Jet Components in Quasars}
\begin{figure}
    \resizebox{\hsize}{!}{\includegraphics{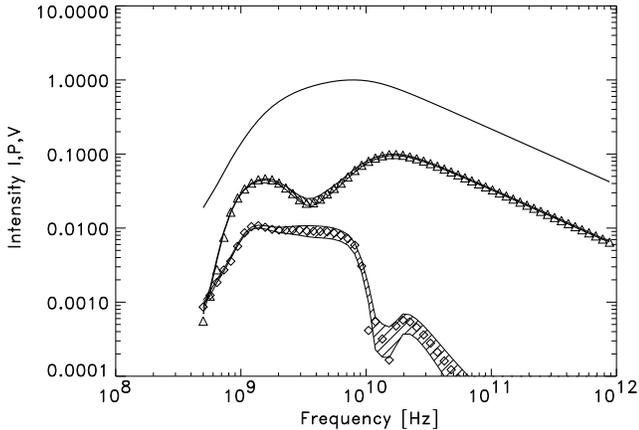}} 
  \caption{The spectrum of a typical inner jet component in total intensity
  I (solid line), LP (triangles), CP (diamonds). The fluxes are normalized to
  the peak flux in I. Energy equipartition between electrons and $B$-field
  is assumed. The electron spectral index is $s =2.5$ and the power-law
  extends from $\gamma_{\mathrm{min}} = 20$ to $\gamma_{\mathrm{max}} = 10^4$.
  The component is not homogeneous but is part of a jet with
  $z_{\mathrm{out}}/z_{\mathrm{in}} = 10$ and a tight helical field
  ($B_z/B_\phi = 1/3$) seen under an angle of $55^\circ$ in the 
  rest frame of the radiating plasma.\label{JetComp} } 
\end{figure}
In the bright jet sources 3C\,84 and 3C\,273 (Homan \& Wardle
\cite{HW99}) it has been demonstrated, that CP can be detected in the
core and the innermost jet component. The degree of LP is equal or
less than CP in the inverted spectrum cores.  Various other components
show 0.5\% circular polarization in these sources.  For 3C 273 it is
claimed, that the circular polarization is predominately associated
with new ejected jet components.  This has to be taken with caution as
Taylor (\cite{T98}) reports a large rotation measure of $RM =
-1900\,$rad\,m$^{-2}$ for the core of 3C\,273 and almost equally large
$RM$ for the core of 3C\,279.  The reduced linear polarization in
3C\,273 may therefore be due to depolarization in surrounding cold
gas.  They also report that the component CW in 3C\,279 is +1.2\%
circularly polarized, while the linear polarization of component CW is
13\% on average. A model for a typical jet component is shown in
Fig.\,\ref{JetComp}. In contrast to Sgr A* we have here less reduction
of polarization due to turbulence (as we only look at one spatially
resolved jet component) and less Faraday depolarization. The model
invokes a higher $\gamma_{\rm min}$ of order 20 and a power law up to
$\gamma_{\rm max}=10^4$ that produces emission well into the optically
thin regime, which is known not to exist in Sgr A*. This gives one
the characteristic LP-to-CP ratio of $\sim10$ observed in quasars and
recently also in X-ray binaries (Fender et al. \cite{Fender2002}).  

It is also interesting to note that we can produce the observed CP
with a $\gamma_{\rm min}$ of order 20. This is somewhat higher than
the rather low values found by Wardle et al. (\cite{Wardle1998}) and does
not place quite so stringent constraints on the energy budget and the
matter content of the jet. Somewhat more realistic energy
distributions at low-energies other than a sharply cut-off power-law
may further relief these constraints.

\section{Polarization Variability} \label{polvar}
In the presence of a turbulent magnetic field the degree of
polarization (both circular and linear) and position angle will vary
stochastically with a timescale $\Delta t = \Delta z/v_A$.
Polarization variability is expected
to be faster than variations in total flux, because the relevant length-scale
is the outer turbulent length-scale $k_{\mathrm{out}}^{-1}$.
With $v_A \approx c_S \approx
c/\sqrt{3}$ we get a characteristic variability time
\begin{equation} \label{vart}
  \Delta t = \frac{\sqrt{3}\,R}{c (k_{\mathrm{out}} R)}
  \propto \nu^{-(s+4)/(s+4+4a/3)}
\end{equation}
The effective radius at a given frequency can be determined from a model
for the total flux, while the turbulent scale can be derived from polarization
of optically thin emission or estimated from the ratio
$\left<\pi_L\right>/\left<\pi_C\right>$.
Turbulence also leads to decorrelation of polarization and position angle
across frequency bands. Polarization properties from optical surfaces
a distance $k_{\mathrm{out}}^{-1}$ apart should be uncorrelated.
This translates into
a relative distance $\Delta z/z = \sin\theta/(k_{\mathrm{out}} R)$
and both LP and CP
should vary across frequency bands $\Delta \nu$ set by the
turbulence in the source
\begin{equation} \label{varf}
  \frac{\Delta \nu}{\nu} = \left(1+\frac{a/3}{s/4+1}\right)\;
  \frac{\sin\theta}{k_{\mathrm{out}} R}
\end{equation}
The suggested model for Sgr A$^*$ with the Alfven velocity of
$v_A = 2\,10^9\,$cm/s and $k_{\mathrm{out}} R = 24$ implies a
variability timescale of $\Delta t \approx
16$\,h at 1\,GHz and an accordingly shorter variability time scale at higher
frequencies. The decorrelation across frequency bands is expected at a width
$\Delta \nu/\nu \approx 2.4\cdot 10^{-2}$. Longer integration times than set by
Eq. (\ref{vart}) and frequency bands wider than given by Eq. (\ref{varf}) will
produce smaller polarization measurements than intrinsically available in the
source. 
\section{Discussion} \label{discus}

Recent observations of radio circular polarization in AGN, X-ray
binaries, and the Galactic Centre black hole suggest that CP at the
0.3\%-1\%-level is common to many self-absorbed synchrotron
sources. Faraday rotation and conversion in a magnetized and therefore
bi-refringent plasma produce enhanced circular and reduced linear
polarization.  Both processes are sensitive to the presence of
low-energy electrons and the orientation of the global magnetic field.

The standard jet model for compact radio cores with a helical plus a
turbulent magnetic field can well reproduce the circular and linear
polarization spectrum of sources like Sgr A* and M81* with their high
CP-to-LP ratio. The suppression of LP is achieved by the presence of a
significant number of low-energy electrons in the source and an
absence of an optically thin power-law extending to higher
energies. The same model can also explain the typical level of
circular polarization in blazars and the CP-to-LP ratio observed in
blazars and X-ray binary jets. In this case a number of low-energy
electrons is reduced with respect to the Sgr A* model and a power-law
in the electron distribution exists.

For Sgr A* the number of low-energy electrons producing conversion and
depolarization needs to be significantly higher (by 2-3 orders of
magnitude) than the number of radiating hot electrons.  This means
that a large fraction of the outflowing jet material is in the form of
hidden matter shielded by self-absorption. This increases the estimates
of the total jet power, which can be 5 orders of magnitude higher than
the synchrotron luminosity. If one presumes that this power has to be
provided by an accretion flow, the minimum accretion rates of
$10^{-9..-8} M_{\odot}$/yr, previously estimated from
``maximally-efficient'' jet models for Sgr A* (Falcke et
al.~\cite{FMB93}; Falcke \& Biermann \cite{FB99}) need to be raised to
about $10^{-6} M_{\odot}/$yr. This is quite consistent with recent
estimates of Bondi-Hoyle accretion rates onto Sgr A* (Baganoff et
al. \cite{Baganoff2002}) and with suggestions for a coupled jet plus ADAF
model (Yuan et al. \cite{Yuan2002}), where the emission from the
accretion disk is highly suppressed with respect to the jet.

It is also interesting to note that to fit the CP with conversion one
requires an asymmetry in the magnetic field components.  This is
naturally achieved by a helical magnetic field as is presumed to exist
in jets. A symmetric configuration, e.g. a tightly wound helix or even
a toroidal magnetic field structure would have difficulties to produce
the observed level of CP.

The stable handedness of CP over 20 years also implies a long-term
stable component of the unidirectional field along the line-of-sight.
This indicates that the polarity of the magnetic field (the ``magnetic
north pole'') has remained constant over the last two decades.  In
view of the rather short accretion time scale in Sgr A$^*$ one could
also speculate that this polarity is related to the accretion of a
stable large-scale magnetic field which is accreted and expelled via
the jet. The same can be said about blazars and X-ray binaries, where
the stability found in GRS1915+105 by Fender et al. (\cite{Fender2002})
is particularly interesting since the intrinsic accretion time scales
in X-ray binaries are much shorter than those in supermassive black
holes.

Another important aspect of CP measurements is the question of the
matter content of jets. We find that the constraints from CP of
individual jet components for the jet power in blazars are not quite
as severe as previously claimed and a statement in support of a pure
electron/positron jet has to viewed with caution. For Sgr A* or M81*
the situation may be different. If the depolarization is indeed
intrinsic to the jet and not a surrounding medium (Agol \cite{Agol2000},
Quataert \& Gruzinov \cite{Quataert2000}),
one needs a high Faraday optical depth in the
source, which can only be achieved by an excess of ``warm''
($1\la\gamma\la100$) electrons in an electron/proton plasma. 

While we have here assumed that all electrons are distributed in a
single power-law, the actual situation may be quite different. For Sgr
A* a power-law is actually not needed and we could obtain rather
similar results with a two-temperature electron distribution, with
temperatures corresponding to $\gamma_{\rm min}$ and $\gamma_{\rm
max}$ respectively. This is not quite possible in blazars or bright
X-ray binary jets, where extended electron power-laws are directly
observed in the optically thin regime. It could well be that the
radiative inefficiency of Sgr A* is due to the lack of effective shock
acceleration that would increase the number of high-energy electrons
with respect to the number of low-energy electrons (and in turn
decrease the CP-to-LP ratio). The origin of these different electron
distributions and their role for the radio-loudness of jet sources
should be a very exciting question for further research.

By improving our sensitivity and imaging all four Stokes parameters at
multiple frequencies in the future, it will be possible to construct
models of the entire emission and transfer processes in the
source and determine the composition and energy spectrum of the
relativistic plasma within jets.

\appendix
\section{The Radiative Transfer Problem}\label{AppA}
In a weakly anisotropic medium the propagation of electromagnetic radiation
is determined by the dielectric tensor $\epsilon_{ij}$. The normal modes in a
magneto-active, anisotropic plasma are quasi-transverse, but they are not
orthogonal. The transfer of the intensity tensor $I_{ij}$ along a ray
path\footnote{In this section \ref{AppA} the path length along the ray
is called $s$ not be confused with the spectral index of a
power-law distribution.} 
$\vec{k}/k$ is given by (Sazonov \cite{S69}, Zheleznyakov et al. \cite{Z74})
the emissivity $S_{ij}$ and $\epsilon_{ij}$\
\begin{equation} \label{transEq}
   \frac{{\mathrm{d}} I_{ij}}{{\mathrm{d}} s} = S_{ij} + i\frac{\omega}{c}
\left(\epsilon_{ik} I_{kj} - I_{ik} (\epsilon^{\dagger})_{kj}\right) \, .
\end{equation}
For transverse waves the intensity tensor is a $2\times 2$ tensor perpendicular
to the wave vector $\vec{k}$ and for the emissivities and the dielectric
tensor only the transverse components enter  Eq. (\ref{transEq}). The
$\dagger$ indicates the complex conjugate and transpose of $\epsilon$.

In a magneto-active plasma with the local magnetic field $B$ unperturbed by the
wave along the $z$-axis the dielectric tensor has only one symmetry
$\epsilon_{xy} = - \epsilon_{yx}$.
The tensor $\epsilon$ may be separated into a hermitian $(H)$ and
anti-hermitian $(A)$ part according to
\begin{eqnarray}
  \epsilon_{ij} & = & \epsilon_{ij}^{(H)} + \epsilon_{ij}^{(A)} \\
  \epsilon_{ij}^{(H)} & = & \frac{1}{2}\left( \epsilon_{ij}
  + \epsilon_{ij}^*\right) \\
  \epsilon_{ij}^{(A)} & = & \frac{1}{2}\left( \epsilon_{ij}
  - \epsilon_{ij}^*\right) \quad .
\end{eqnarray}
The hermitian $\epsilon^{(H)}$ describes absorption processes, while the
action of $\epsilon^{(A)}$ in Eq. (\ref{transEq}) conserves the total
intensity and rotates the polarization vector of elliptically polarized
radiation on the Poincare sphere,
which is formed by the normalized Stokes parameters $(Q,U,V)/I$
(Kennet \& Melrose \cite{KM98}). This generalized rotation consists of Faraday
rotation and ordinary conversion between $U$ and $V$ and extraordinary
conversion between $Q$ and $V$. From Eq. (\ref{transEq}) the transfer
of the Stokes parameters for polarized waves follow
\begin{eqnarray}\label{TransPro}
  \frac{\mathrm{d} I}{\mathrm{d} s} & = & \eta_I - \kappa_I I - \kappa_Q Q  
  -\kappa_V V -\kappa_U U \\
  \frac{\mathrm{d} Q}{\mathrm{d} s} & = & \eta_Q - \kappa_I Q
  - \kappa_Q I  - \kappa_F U -h_Q V \\
  \frac{\mathrm{d} U}{\mathrm{d} s} & = & \eta_U - \kappa_I U
  - \kappa_U I + \kappa_F Q - \kappa_C V   \\ \label{TransProEnd}
  \frac{\mathrm{d} V}{\mathrm{d} s} & = & \eta_V - \kappa_I V  -\kappa_V I 
  + h_Q Q + \kappa_C U \quad .
\end{eqnarray}
The transport is described by the transport coefficients for absorption 
\begin{eqnarray}
  \kappa_I & = & \frac{\omega}{2 c}\left\{ \Im (\epsilon_{yy}) + \cos^2\theta\,
  \Im (\epsilon_{xx}) 
   + \sin^2\theta\, \Im (\epsilon_{zz}) \right. \nonumber \\ & & \left.
  - \sin\theta\, \cos\theta\, (\Im (\epsilon_{zx}) 
  + \Im(\epsilon_{xz}))\right\}
\end{eqnarray}
\begin{eqnarray}
  \kappa_Q  & = & \frac{\omega}{2 c}\left\{ -\Im(\epsilon_{yy}) 
  + \cos^2\theta\, \Im(\epsilon_{xx}) + \sin^2\theta\, \Im(\epsilon_{zz})
  \right. \nonumber \\ & & \left.
  - \sin\theta\, \cos\theta\, (\Im (\epsilon_{zx}) 
  + \Im(\epsilon_{xz}))\right\}
\end{eqnarray}
\begin{equation}
  \kappa_V = -\frac{i \omega}{2 c}\left\{ 2\cos\theta\, \Re(\epsilon_{xy}) 
  + \sin\theta\, (\Re(\epsilon_{yz}) - \Re(\epsilon_{zy}))\right\}
\end{equation}
\begin{equation}
  \kappa_U = -\frac{\omega}{2 c} 
  \sin\theta\,(\Im (\epsilon_{yz}) + \Im(\epsilon_{zy})) \quad .
\end{equation}
The rotation coefficients for Faraday rotation $\kappa_F$ and
conversion $\kappa_C$, $h_Q$
are 
\begin{equation} \label{FRotprinz}
  \kappa_F = \frac{\omega}{2 c}\left\{ 2\cos\theta\, \Im(\epsilon_{xy}) 
  + \sin\theta\, (\Im(\epsilon_{yz}) - \Im(\epsilon_{zy}))\right\}
\end{equation}
\begin{eqnarray}
  \kappa_C  & = &  -\frac{\omega}{2 c}\left\{ \Re(\epsilon_{yy})
  - \cos^2\theta\, \Re(\epsilon_{xx}) - \sin^2\theta\, \Re(\epsilon_{zz})
  \right. \nonumber \\ & & \left.
  - \sin\theta\, \cos\theta\, (\Re (\epsilon_{zx}) 
  + \Re(\epsilon_{xz}))\right\}
\end{eqnarray}
\begin{equation}
  h_Q = \frac{\omega}{2 c} \sin\theta\, (\Re(\epsilon_{yz})
  + \Re(\epsilon_{zy})) \, .
\end{equation}
In the case of an isotropic distribution of unperturbed particles $f(p)$
two additional symmetries of the dielectric tensor $\epsilon$ appear
$\epsilon_{yz} = -\epsilon_{zy}$ and $\epsilon_{xz} = \epsilon_{zx}$.
Therefore the emissivity and absorption coefficient for $U$ and the
extraordinary conversion $h_Q$ vanish.

\section{Dielectric Tensor and Plasma Kinetic Theory}\label{AppB}
The reaction of a distribution of charged particles $f_0(\vec{p})$
in a magnetic field $\vec{B}$ to a perturbing wave can be
derived from the linearized first-order perturbation to the
Vlasov equation (Montgomery \& Tidman \cite{MT64}).
In Fourier-Laplace space the perturbation of the particle distribution
$f_1$ is given in terms of a propagator  $G(\phi')$
\begin{equation} \label{perturbf}
  f_1 = \frac{e}{\Omega} \int_{\pm\infty}^\phi {\mathrm d}\phi'\,
  G(\phi')\, 
  \left(\tilde{E} - i\frac{v'}{s}\wedge (k \wedge \tilde{E})\right)
  \frac{\partial f_0(\vec{p})}{\partial p'} \, .
\end{equation}
Here $\tilde{E}$ is the perturbed electric field of the wave, $e$ the charge
and $v'$ the velocity of particles characterized by a phase $\phi'$ along
their path. $s = s_0 - i\omega$ is a complex frequency and
$\Omega = \Omega_0/\gamma$ the gyro-frequency of particles with energy
$\gamma m c^2$. The sign of the lower boundary in Eq. (\ref{perturbf}) is
determined by the charge of the particles ($-\infty$ for electrons; $+\infty$
for positrons). The propagator is 
\begin{equation}
   G(\phi')  = \exp\left[\int_{\phi'}^{\phi} {\mathrm d} \phi''
   (s + i k\cdot v'')/\Omega \right] \, ,
\end{equation}
which can be used to calculate the Fourier-Laplace transform of the
current density in response to the perturbing wave
\begin{equation} \label{current}
   \tilde{j} = e n_0 \int {\mathrm d}^3 p\, f_1(\vec{p})\, \vec{v} \, ,
\end{equation}
where $n_0$ is the particle density in the plasma. The components
of the dielectric tensor can then be read off the relation
\begin{equation}
  \epsilon_{ij} \tilde{E}_j =  4 \pi\, i\, \omega^{-1}\, \tilde{j}_i  \, ,
\end{equation}
For isotropic distributions $f_0(p)$ the second term in parentheses in
Eq. (\ref{perturbf}) vanishes and the calculation of $\epsilon_{ij}$ is
greatly simplified.
The $\phi'$-integration in Eq. (\ref{perturbf}) results in a factor
$-i\Omega/(n \Omega + k_\parallel v_\parallel - \omega - i\tilde{\epsilon})$,
where $\tilde{\epsilon}$ is small and positive, because no emission process
is described by the dielectric tensor. In the limit $\tilde{\epsilon}
\rightarrow 0_+$ the momentum integrals in Eq. (\ref{current}) enclose
singularities and must be interpreted as
\begin{equation} \label{prinzipal}
  \lim_{\tilde{\epsilon}
  \rightarrow 0_+} \int {\mathrm{d}} x\,\frac{\tilde{f}(x)}
  {x-(z + i \tilde{\epsilon})}
  = {\cal{P}}\!\int {\mathrm{d}} x\,\frac{\tilde{f}(x)}{x-z}
  + \pi i \tilde{f}(z)\, ,
\end{equation}
where ${\cal{P}}\!\int$ indicates a principal value integral.
The terms containing $\delta$-functions lead to the imaginary part of the
diagonal elements of $\epsilon_{ij}$ and to the real part of the off-diagonal
elements. They describe absorption of the four Stokes parameters. The terms
connected with principal value integrals are $90^\circ$ off in the
complex $\epsilon_{ij}$-plane and describe the generalized rotation.

\section{Faraday Rotation and Conversion}\label{AppC}
With the help of Eq. (\ref{prinzipal}) in Eq. (\ref{perturbf}) we get for
Faraday rotation (\ref{FRotprinz}) by an isotropic distribution of particles
\begin{eqnarray} 
  \kappa_F  & = &  4 \pi \cos \theta \frac{\omega_p^2}{c \Omega_0}
  \sum_{n=1}^{\infty} {\cal{P}}\!\int {\mathrm{d}} p_\parallel
  \int {\mathrm{d}} p_\perp\, 
  \frac{\partial f_0(p)}{\partial p_\perp} \nonumber \\
  & & \times \frac{p_\perp^2 n \Omega^2}
  {n^2 \Omega^2 - (k_\parallel v_\parallel - \omega)^2}
  \frac{n \Omega}{k_\perp v_\perp}\, J_n(z)\, J'_n (z) \, .
\end{eqnarray}
In general Faraday rotation is represented by an series of Bessel-functions
$J_n(z)$ of order $n$ and their derivatives. The argument of the
Bessel-functions is $z = k_\perp v_\perp/\Omega$.
In the high-frequency cold plasma limit only the $n=1$ term is important
and the momentum integral can be solved by partial integration to get
the classical limit
\begin{equation} 
  \kappa_F^{(c)}  =  \frac{\omega_p^2 \Omega_0}{c \omega^2}\cos\theta \;. 
\end{equation}
In the ultra-relativistic limit $\gamma \gg 1$ Sazonov (1969) derived for
power-law distributions in energy with $s > -1$
the Faraday rotation coefficient
\begin{equation} 
  \kappa_F^{(r)}  =  \frac{\omega_p^2 \Omega_0}{c \omega^2}\,
  \frac{(s+2)}{(s+1)}
  \,\frac{ \ln\gamma_{\mathrm{min}}}{\gamma_{\mathrm{min}}^{s+1}}
  \cos\theta\;, 
\end{equation}
where the plasma frequency $\omega_P$ must be taken for the density of
particles under consideration.
In the high-frequency limit for a cold plasma the conversion coefficient
$\kappa_C$ is
\begin{equation} 
  \kappa_C^{(c)} = -\frac{\omega_p^2 \Omega_0^2}{2 c \omega^3}\sin^2\theta
  \;. 
\end{equation}
Again Sazonov (\cite{S69}) gave an ultra-relativistic approximation for
conversion by a power-law distribution of particles
\begin{equation} 
  \kappa_C^{(r)}  =  \kappa_{C (c)}
  \frac{2}{s-2}\left(\gamma_{\mathrm{min}}^{-(s-2)}
  -\left(\frac{\omega}{\Omega_0 \sin\theta }
  \right)^{-(s-2)/2}\right)\; . 
\end{equation}

\section{Transport Coefficients}\label{AppD}
Here we summarise the transport coefficients for a relativistic plasma
with a normalized power-law distribution $N(\gamma)$d$\gamma =
n_e f_0(\gamma)$d$\gamma$ above a lower cut-off energy $\gamma_{\mathrm{min}}
m_e c^2$
\begin{equation} \label{powerl}
   f_0(\gamma) = \frac{s-1}{\gamma_{\mathrm{min}}^{1-s}}\gamma^{-s}\, .
\end{equation}
All the absorption and rotation coefficients can be scaled to the inverse
length-scale
\begin{equation}
  l_0^{-1} = \kappa_0 = \pi \frac{\nu_p^2}{c}\frac{\nu_B}{\nu^2} = r_e c n_e
  \frac{\nu_B}{\nu^2} \, ,
\end{equation}
with $r_e$ the classical electron radius, $n_e$ the particle density,
$\nu_B = e B/( 2 \pi mc)$ the cyclotron frequency  and
$\nu_p = \sqrt{n_e e^2 /(\pi m_e)}$ the plasma frequency.
The transport coefficient for Faraday rotation by power-law electrons
(Eq. \ref{powerl}) is
\begin{equation}\label{Frotate}
  \kappa_F = 2  \kappa_0 \cos\theta\, \frac{s+2}{s+1} \,
  \frac{ \ln\gamma_{\mathrm{min}}}{\gamma_{\mathrm{min}}^{s+1}}
\end{equation}
and for conversion 
\begin{equation}
  \kappa_C =  -\kappa_0 \frac{\nu_B}{\nu} \sin^2\theta
  \frac{2}{s-2}\left(\frac{1}{\gamma_{\mathrm{min}}^{(s-2)}}
  -\left(\frac{\nu_B}{\nu}
  \right)^{(s-2)/2}\right)\, .
\end{equation}
The three absorption coefficients are
\begin{eqnarray} 
   \kappa_I & = & \kappa_0 \sin\theta
   \left(\frac{\nu_{B\perp}}{\nu}\right)^{s/2} \nonumber \\ & & \times
   \frac{3^{(s+1)/2}}{4}\Gamma\left(\frac{s}{4}
   +\frac{11}{6}\right) \Gamma\left(\frac{s}{4}+\frac{1}{6}\right) \, ,
\end{eqnarray}
\begin{equation} 
   \kappa_Q  = 
   \frac{s+1}{s+7/3} \kappa_I \, ,
\end{equation} 
\begin{eqnarray}
   \kappa_V & = & \kappa_0 \cos\theta
   \left(\frac{\nu_{B\perp}}{\nu}\right)^{(s+1)/2}
   \frac{(s+3)(s+2)}{s+1)}\nonumber \\ & & \times    \frac{3^{(s+1)/2}}{4}
    \Gamma\left(\frac{s}{4}
   +\frac{11}{12}\right) \Gamma\left(\frac{s}{4}+\frac{7}{12}\right) \, .
\end{eqnarray}
The emission coefficients can be scaled to the
emissivity 
\begin{equation}
  \eta_0 = \pi \frac{\nu_p^2}{c^3}\,\nu_B\, m_e c^2 =
  (r_e/c)\, n_e\,\nu_B\, m_e c^2
\end{equation}
and for the power-law distribution of Eq. (\ref{powerl}) we get 
\begin{eqnarray} 
   \eta_I & = & \eta_0 \sin\theta
   \left(\frac{\nu_{B \perp}}{\nu}\right)^{(s-1)/2}
   \nonumber \\ & & \times
   \frac{3^{s/2}}{2\,(s+1)} \Gamma\left(\frac{s}{4}
   +\frac{19}{12}\right) \Gamma\left(\frac{s}{4}-\frac{1}{12}\right) \, , 
\end{eqnarray}
\begin{equation} 
  \eta_Q =  \frac{s+1}{s+7/3} \eta_I\,; \quad \eta_U = 0 \, ,
\end{equation} 
\begin{eqnarray} 
   \eta_V & = & - \eta_0 \cos\theta
   \left(\frac{\nu_{B \perp}}{\nu}\right)^{s/2}
   \nonumber \\ & & \times\frac{3^{(s-1)/2}\,(s+2)}{2\,s}
   \Gamma\left(\frac{s}{4}
   +\frac{2}{3}\right) \Gamma\left(\frac{s}{4}+\frac{1}{3}\right) \, .
\end{eqnarray}

\end{document}